\pdfoutput=1
\documentclass[aps,pre, twocolumn, groupedaddress]{revtex4-1}
\usepackage{lipsum}
\usepackage{mathtools}
\usepackage{graphicx}
\usepackage{dcolumn}
\usepackage{amsmath}    
\usepackage{amssymb}
\usepackage{bm}
\usepackage{hyperref}
\usepackage{latexsym}
\usepackage{verbatim}
\usepackage[normalem]{ulem}
\usepackage{color}
\usepackage[caption=false]{subfig}
\def\beq{\begin{equation}}
\def\eeq{\end{equation}}

\def\beq{\begin{equation}}                          
\def\eeq{\end{equation}}                          
\def\bea{\begin{eqnarray}}                          
\def\eea{\end{eqnarray}}

\DeclareRobustCommand{\uvec}[1]{{%
  \ifcsname uvec#1\endcsname
     \csname uvec#1\endcsname
   \else
    \bm{\hat{\mathbf{#1}}}%
   \fi
}}
\preprint{}
\preprint{}
\begin{document}
\title{Effective single component description of steady state structures of passive particles in an active bath}
\author{Jay Prakash Singh$^{1}$}
\email{jayps.rs.phy16@itbhu.ac.in}
\author{Sudipta Pattanayak$^{2}$}
\email{pattanayak.sudipta@gmail.com}
\author{Shradha Mishra$^{1}$}
\email{smishra.phy@iitbhu.ac.in}
\affiliation{$^{1}$Department of Physics, Indian Institute of Technology (BHU), Varanasi, India 221005}
\author{Jaydeb Chakrabarti$^{2}$}
\email{jaydeb@bose.res.in}
\affiliation{$^{2,}$S. N. Bose National Centre for Basic Sciences, J D Block, Sector III, Salt Lake City, Kolkata 700106}
\date{\today}
\begin{abstract}
We model a  binary mixture of passive  and active Brownian particles in two dimensions using the effective interaction between passive particles in the active bath. The activity of active particles and the size ratio of two types of particles are two control parameters in the system. The effective interaction is calculated from the average force on two particles generated by the active particles. The effective interaction can be attractive or repulsive,  depending on the system parameters. The passive particles form four distinct structural orders  for different system parameters viz; disorder  $(D)$, disordered cluster $(DC)$, ordered cluster $(OC)$, and polycrystalline order $(PC)$. The change in structure is dictated by the change in nature of the effective interaction. We further confirm the four structures using full microscopic simulation of active and passive mixture. Our study is useful to understand the different collective behaviour in non-equilibrium systems. 
\end{abstract}
\maketitle
\section{Introduction \label{introduction}}

A complex system in general has a host of degrees of freedom out of which only a finite subset could be of interest to describe  certain properties of the system. Such a reduced description, also known as the effective free energy description in terms of a set of selected degrees of freedom while integrating the remaining degrees of freedom of the system, is a well-established technique in equilibrium\cite{likos}. Depletion forces between hard sphere colloids in a dispersion belong to this category, for instance\cite{depletion}. Since many of the equilibrium techniques, including lack of  a free energy based description, break down  for a system out of equilibrium, a reduced description of such systems in analogy to equilibrium is not obvious. 

Non-equilibrium systems exhibit a variety of collective behaviours, for which no reduced description has been attempted so far. A collection of active or self-propelled particles  \cite{Feder,Jp1,Jp2,Rauch} is of current interest where the system is driven out of equilibrium. Examples of active systems range from small scale
of the order of intracellular to macroscopic length scale of few meters \cite{Feder,Tonertu1,Ramaswami,Marchetti,Harada,Bado,Surrey1,Jacob,Mcapp,Helbing1,
Helbing2,Ekku,Hubbard1,Schaller1,Sumino1,Peruani1,Cohen1}, exhibiting a host of non-equilibrium phenomena, like pattern formation \cite{Marchetti}, non-equilibirum phase transition \cite{Vicsek1}, 
large density fluctuations \cite{Chate1,Chate2,Bhatt1,Narayan1}, 
enhanced dynamics \cite{Bechi1,Angel1,Harder1,Sameer1,Viv1,Pablo,Sudipta1,Vive,Bhaskaran1}, motility induced phase separation \cite{Butti1,Cates1,Sese1,Sham1,Solon1} and so on.
Collection of spherically symmetric active Brownian particles (ABP), like the Janus particles, or active colloids show motility induced phase separation (MIPS) at packing density much lower than 
their equilibrium counterparts\cite{athermal}. Recently, MIPS has been extensively studied  in various 
experiments on synthetic colloids, bacterial and cell suspensions \cite{Butti1,Cates1,Sese1,Solon1}.
Motivated by the  MIPS in pure active systems, the mixture of passive and active particles are also explored in various 
experiments and theoretical studies \cite{Bechi1,Marchetti}.  In recent years a number of 
studies are performed to study the phase separation of passive particles on varying system parameters, like activity
of the medium and size of passive particles \cite{Pritha1,Joakim1,Amitdas1,Wang1}.
Phase separation of passive particles in
the mixture can be a potential model to explore the 
clustering and aggregation of big macro-molecules in cellular environment \cite{phase1,phase2,phase3,phase4,phase5,phase6,phase7}. 

As a prototype of colective behaviour in systems out of equilibrium, it is interesting to develop a reduced description of the MIPS. In equilibrium the effective interaction among the solute particles in a bath of solvent particles is measured in terms of the pair correlation function of solvent particles 
$g_2(r)$  \citep{Pritha1,Angel22,Jharder11,Kraf1} or from the average force acting by the solvent particles on two solute particles at a fixed separation in presence \cite{AO1,AO2,AO3,AO4,AO5,jaydebsir1,jaydebsir2}.  There are recent reports on non-equilibrium depletion forces \cite{prlforce,Small,Mallory,Ray,Cohen,Leite,Yamchi,Duzgun,Hua,Baek}. It has been shown in Ref.\cite{prlforce} that
the effective force between passive particles in an active medium can not be accounted for in terms of $g_2(r)$. In addition, the solvent mediated average force on two solute particles depends on the manner in which particles are constrained. Both the features suggest departure from the equilibrium scenario.  However, unlike the equilibrium counterpart, it is not established yet to the best of 
our knowledge how far the effective interaction in these out of equilibrium systems can describe the collective behaviour in steady-state  \cite{Pritha1}.

In this study, we focus on the steady state configuration of the passive particles in the presence of ABPs in terms of the effective potential between two passive particles in the bath. We fix the passive particles at a given separation, while the ABPs perform their motions. The component of the forces between the active and passive particles along the separation vector of the fixed passive particles is computed and averaged over many steady state configurations to give the effective interaction between the passive particles. We perform Brownian dynamics simulations with a large number of passive particles with the effective potential to determine different steady state structures of the 
passive system. We are thus left with an effective single component system of the passive particles alone without the ABPs where the active degrees of freedoms are integrated out into the effective interaction.  We observe four different structural orders  of the passive particles  in parameter space, spanned by  the size ratio 
and activity of passive and active particles: (i) disorder  $(D)$, (ii) disorder-cluster  $(DC)$, (iii) ordered cluster  $(OC)$, and (iv)  polycrystalline  $(PC)$. Finally, observed structures of passive particles are confirmed by full microscopic simulation of a binary mixture of active and passive particles.

In the rest of the paper, we discuss the details of the binary model system in section\ref{model}. Section \ref{CDF} discusses the result of effective force between two passive particles in the presence of small ABPs. In section \ref{FDP} we show the effect of the effective force on a purely passive system and the characteristics of four structures are discussed in detail. Finally, we conclude the paper with a summary and discussion \ref{conclusion} .
\begin{figure}[ht]
\centering
\includegraphics[width=0.90\linewidth]{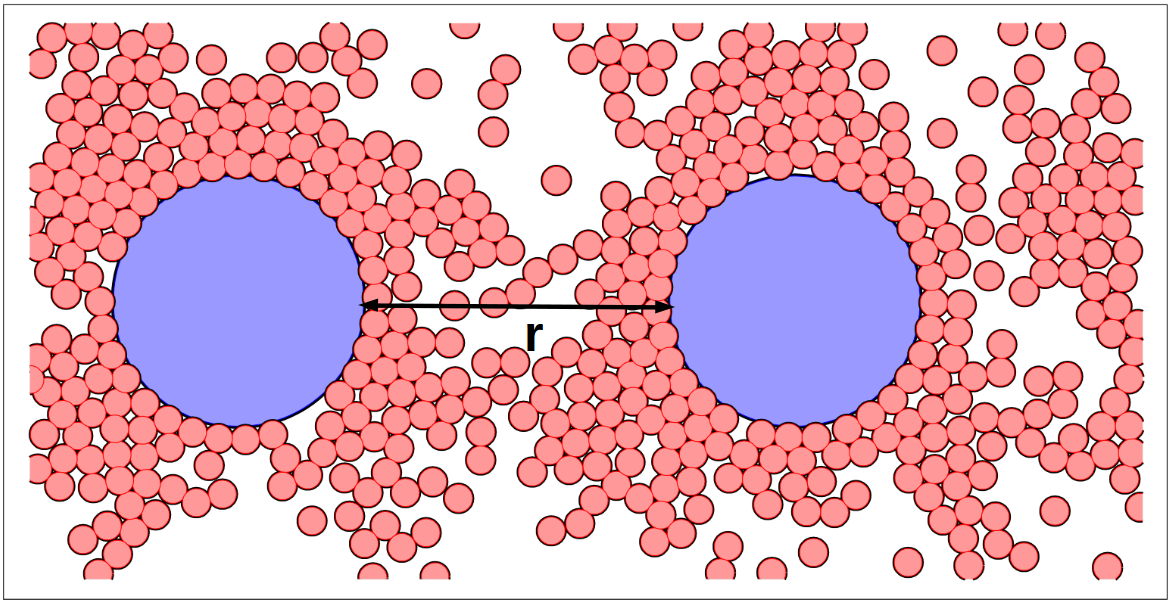}
	 \caption{(color online) Plot shows the model picture of small ABPs and big passive particles with size ratio $S=10$ to calculate the effective potential $V^{eff}(r)$ on passive particle separated by distance `$r$' exerted by the active depletant. Blue and red particles show passive particles and ABPs. The black arrowed line shows the surface to surface distance `r' between two passive particles. Number of ABPs and passive particles are $N_a=1000$ and $N_p=2$ respectively.}
\label{fig:fig1}
\end{figure}
\section{Model \label{model}}
Our system consists of a binary mixture of $N_a$  ABPs of radius $r_a$, and $N_p$ passive particles of  radius $r_p$ moving in two-dimensions $L_x \times L_y$ with the periodic boundary conditions. We define size-ratio of the particles $S = r_{p}/ r_{a}$. 
Let us represent the position vector of the center of the $i^{th}$ ABP and  passive particle by ${\bf r}_{i}^{a}(t)$ and ${\bf r}_{i}^{p}(t)$, respectively at time $t$. The orientation of $i^{th}$ ABP  is represented by a unit vector $ {\bf n}_i(t) = (\cos{\theta_i}(t),\sin{\theta_i}(t))$. The dynamics of the active particle is governed by the overdamped Langevin equation
\begin{equation}
	\partial_t{\bf{r}}_i^{a}=v{\bf{n}_i}+\mu_1\sum_{j\neq i}{\bf {F}}_{ij}
\label{eq(1)}
\end{equation}
 \begin{equation}
	 \partial_t\theta_i(t)={\eta}^r_i(t)
\label{eq(2)}
\end{equation}
The first term on the right hand side (RHS) of Eq. \ref{eq(1)} is due to the activity of the ABPs with active self-propulsion speed  $v$. The rate of change of the orientation $\theta_i$ of the $i^{th}$ ABP is given by Eq.\ref{eq(2)}. The stochastic force ${\bf\eta}^r_i(t)$ at time $t$ is defined as,
$\langle {\bf\eta}^r_i(t){\bf\eta}^r_j(t^{'}) \rangle = 2\nu_r\delta_{ij}\delta(t-t^{'})$. 
$\nu_r$ represents the rotational diffusion constant. 
The persistence length of the ABPs is defined as $l=v/\nu_r$, and the corresponding
persistent time $\tau=1/\nu_r$. We define the dimensionless
activity $\bar{V}=\frac{v}{r_a \nu_r}$. The rotational diffusion constant is kept fixed 
at $\nu_r=0.005$. The size of the active particles $r_a=0.1$, 
The force term ${\bf{F}}_{ij}$ in both equations is due to soft repulsive steric interaction between the particles, ${\bf{F}}_{ij}=-\nabla U(r_{ij})$, where $U(r_{ij})=k(r_{ij}-(r_{\beta i}+r_{\beta'j}))^2$ if $r_{ij}\leq(r_{\beta i}+r_{\beta'j})$ and $r_\beta$, is the radius of active or passive particles for $\beta$ and $\beta'=a$ or $p$ respectively. 
The mobility of both types of particles are kept the same $\mu_1=\mu_2=1.0$ and the force constant $k=1.0$; hence $(\mu_1 k)^{-1}=1.0$ defines the
elastic time scale in the system. The area fraction of the ABPs is $\phi_a = N_a\pi r_a^2/(L_x \times L_y)$ and kept fixed at  $\phi_a=0.5$. The area fraction of passive particles depends on the size of passive particles.
The smallest time step 
considered is $\Delta t=0.001$. The size ratio S and dimensionless activity $\bar{V}$ are two control parameters and they are varied from (1 to 10), and (20 to 160) respectively.

In order to calculate the effective potential between two passive particles we choose $N_p=2$ at positions $\vec{R}_1$ and $\vec{R}_2$ respectively in the sea of ABPs ($N_a=1000$). We keep $\vec{R}_1$  fixed and slowly vary $\vec{R}_2$ in small steps of $\delta x = 1.5 r_a$ starting from the zero surface to surface distance between two passive particles. The active particle coordinates are updated according to the Eqs. \ref{eq(1)} and \ref{eq(2)}.   For each configuration at a given distance between two passive particles the system is allowed to reach the steady state.  Typical time for the steady state $\tau=5\times{10^{7}}$. Further we use the steady state configuration to calculate the force ${\mathcal{F}}^{S,\bar{V}}(r)$ between two-passive particles at a surface to surface separation $r$, such that ${\mathcal{F}}^{S,\bar{V}}(r)={\bf{F}}_{12}(r)+\sum_{i=1}^{N_a}{\bf{F}}_{1i}(r)$. Here ${\bf{F}}_{12}(r)$ is the force due to passive particle $2^{nd}$ on $1^{st}$, and $\sum_{i=1}^{N_a}{\bf{F}}_{1i}(r)$ represents the sum of all the forces due to active particles on $1^{st}$ passive particle for a given configuration of two passive particles at separation r.  Then the potential is calculated by integrating the force over the distance $V^{eff}(r) = \int_{-\infty}^{r}{\bf \mathcal{F}}^{S,\bar{V}}(r)dr$  \cite{jaydebsir1,jaydebsir2}. Here we set the lower limit as half of the box-length. To improve the quality of data, $200$ independent realisations of the similar system is designed. 
Now we define the coarse-grained model to study the system with effective potential $V^{eff}(r)$. Here the system consists of a collection of passive particles only without any ABPs,  interacting with the force $\mathcal{F}^{S, \bar{V}}(r)$ 
calculated from the effective potential. We take  $N_p=400$ in  two dimensions with linear dimensions $L_x=L_y=800r_a$ with the periodic boundary conditions in both the directions. Here the size of the passive particles is kept the same
as for the corresponding potential $V^{eff}(r)$ obtained from the binary system. Hence area fraction of particles $\phi_p$ is different for different ranges of potential used.
The position update of passive particles in the coarse-grained simulations is given by the over-damped Langevin equation
\begin{equation}
	\partial_t{\bf{r}}_i^{p}=-\mu_2\sum_{j\neq i}{\bf {\nabla}}V_{ij}^{eff}(r)+\sqrt{2D_T}\eta^R_{i}(t).
	\label{eq(3)}
\end{equation}
 The first term on the right-hand-side (RHS) of Eq.\ref{eq(3)} defines the effective interaction
force between the passive particles pair $i$ and $j$, where $r$ is the surface to surface separation
between $i^{th}$ and $j^{th}$ passive particles. 
The translational noise $\eta^R_{i}(t)$ at time $t$ is defined as,
$\langle\eta^R_{i}(t)\eta^R_{j}(t^{'})\rangle=\delta_{ij}\delta(t-t^{'})$. $D_T=1.0$ represents the translational diffusion constant. All other parameters are same as defined in section \ref{model}. We consider
total simulation time steps $t=5 \times 10^{6}$. All the physical quantities calculated here are averaged
over 50 realisations of the random noise.  
 Other details are the same as discussed previously. The system is simulated for potentials obtained for the 
 different combination of $S$ and $\bar{V}$.
\begin{figure}[ht]
\centering
\includegraphics[width=0.850\linewidth]{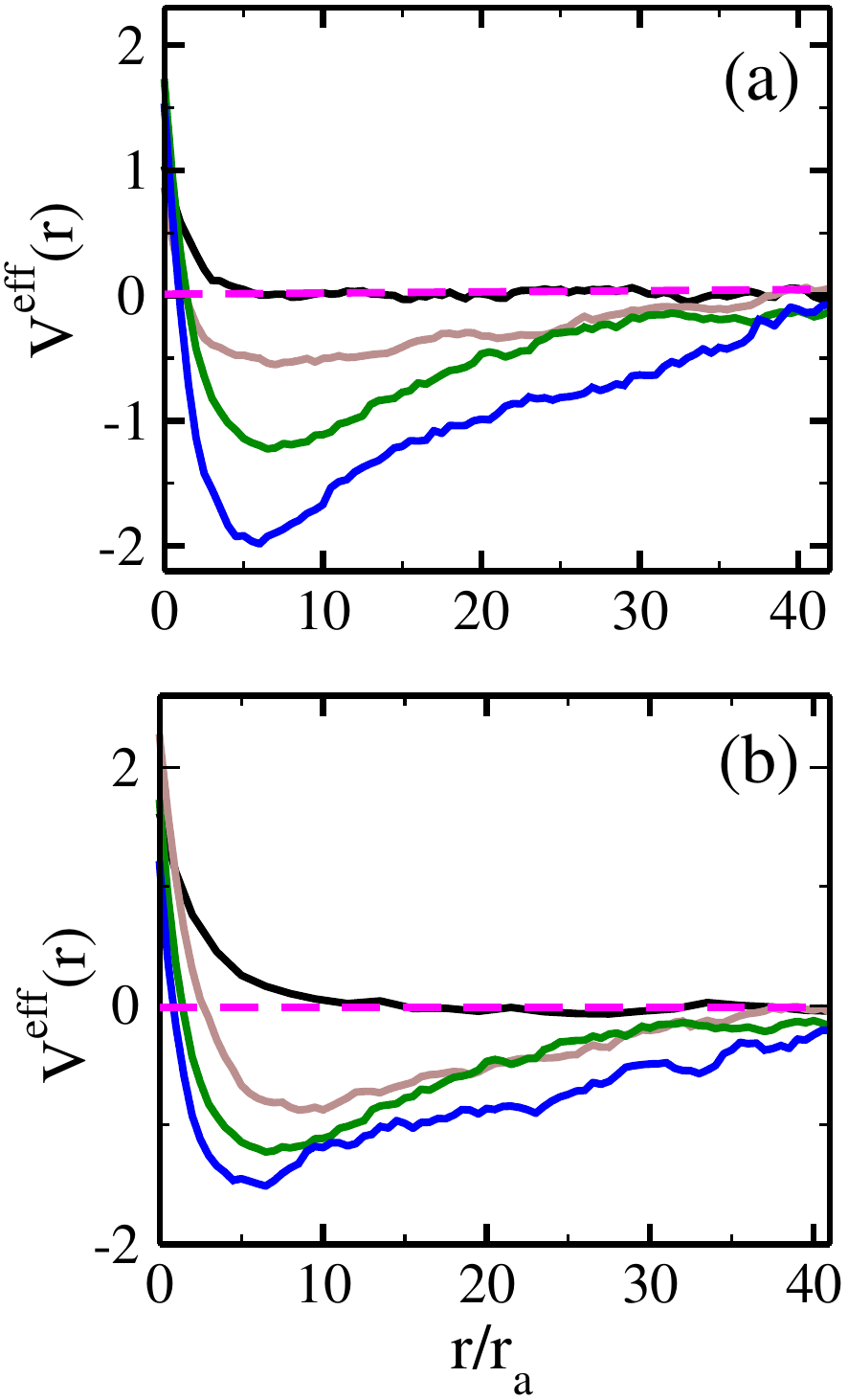}
	 \caption{(color online) {In this panel, we show the variation of numerically obtained active depletant potentials $V^{eff}(r)$ on passive particles for various $\bar{V}$ and size ratio $S$. Where $\frac{r}{r_a}$ is the passive particles separation distance from surface to surface normalise by ABP radius $r_a$. Panel (a): shows the variation of potentials with colors black, brown, green, and blue lines keeping fixed $\bar{V}=160$ for S=4,5,8, and 9. Panel (b): black, brown, green, and blue lines represent the variation of potentials for different $\bar{V}=40,80,120$ and 160, for fixed S=10. The magenta line shows the reference line for mean potential zero. Number of ABPs and passive particles are $N_a=1000$ and $N_p=2$}}
\label{fig:fig2}
\end{figure}
\section{Results and discussions}
\subsection{Effective potential between passive particles\label{CDF}} 
A schematic of the system at a fixed surface-surface separation (r) of passive particles is shown in Fig.\ref{fig:fig1}. 
The results for the effective potential $V^{eff}(r)$ vs. scaled surface-surface distance between two passive particles $\frac{r}{r_a}$ are shown in Fig.\ref{fig:fig2}(a) and (b) for different combinations of activity $\bar{V}$  and size ratio $S$ respectively. Let us first discuss the results in Fig.\ref{fig:fig2}(a). For a fixed $\bar{V}=160$ and  small $S(=4)$, potential is purely repulsive for small $\frac{r}{r_a}$
and then smoothly decay to zero for large $\frac{r}{r_a} \sim 40$. As we increase $S (\>4)$, the potential becomes attractive with minimum at an intermediate distance and approaches to zero value for large distances. Further, the range and the depth of the attractive minima increases with increasing $S$. 
Similarly we show $V^{eff}(r)$ for different $\bar{V}$ at a fixed $S=10$ in Fig.\ref{fig:fig2}(b). For small $\bar{V} =40$, the potential is purely repulsive for small distances $\frac{r}{r_a}$ and then approaches to zero at large distances. 
As we increase $\bar{V}$, potential starts to develop attractive minima at moderate distances and then approaches to zero at larger $\frac{r}{r_a}$. The depth of attractive minima and range of interaction increases on increasing $\bar{V}$.
\begin{figure}[ht]
\centering
\includegraphics[width=1.0\linewidth]{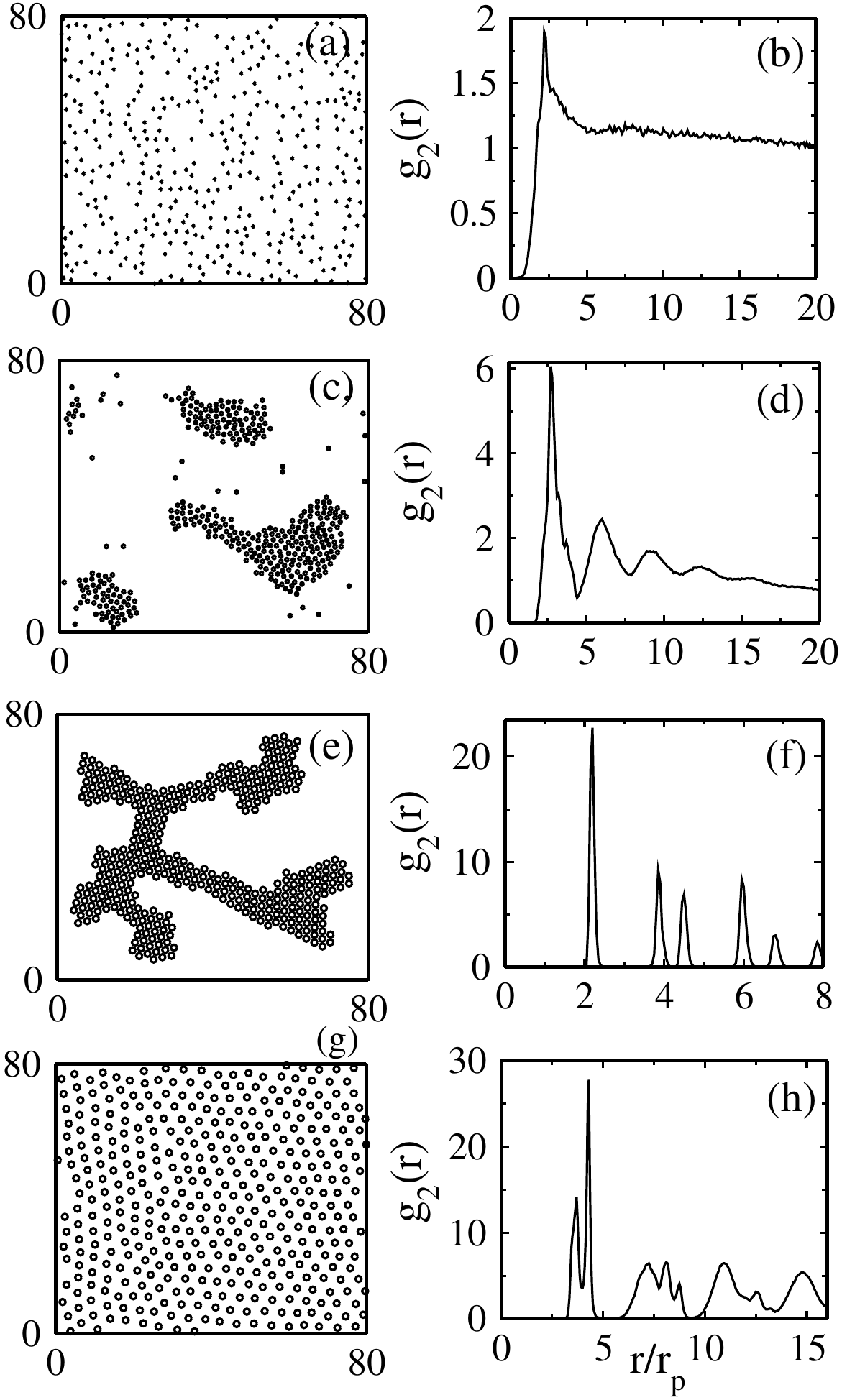}
	 \caption{(color online) {Plots (a), (c), (e), and (g) show the steadystate snapshots for four structures of the passive particles interacting through the potential obtained with different combinations of activity $\bar{V}$ and size ratio $S$ shown in Fig. \ref{fig:fig2}. (a) Disorder structure $(D)$, for $S,\bar{V}=4,160$,
(c) Disorder cluster $(DC)$ , for $S,\bar{V}=5,160$
(e) Ordered cluster $(OC)$ for $S,\bar{V}=10,160$, and 
(g) poly-crystalline structure $(PC)$ for $S,\bar{V}=10,40$.
In the right panel top to bottom (b), (d), (f), and (h), we show the pair correlation distribution $g_2(r)$ vs. $r/r_p$ for same values of $\bar{V}$ and S in the given structures discussed in panels (a), (c), (e), and (g) respectively. Number of passive particles $N_p=400$}}
\label{fig:fig3}
\end{figure} 
\begin{figure}[ht]
\centering
\includegraphics[width=0.950\linewidth]{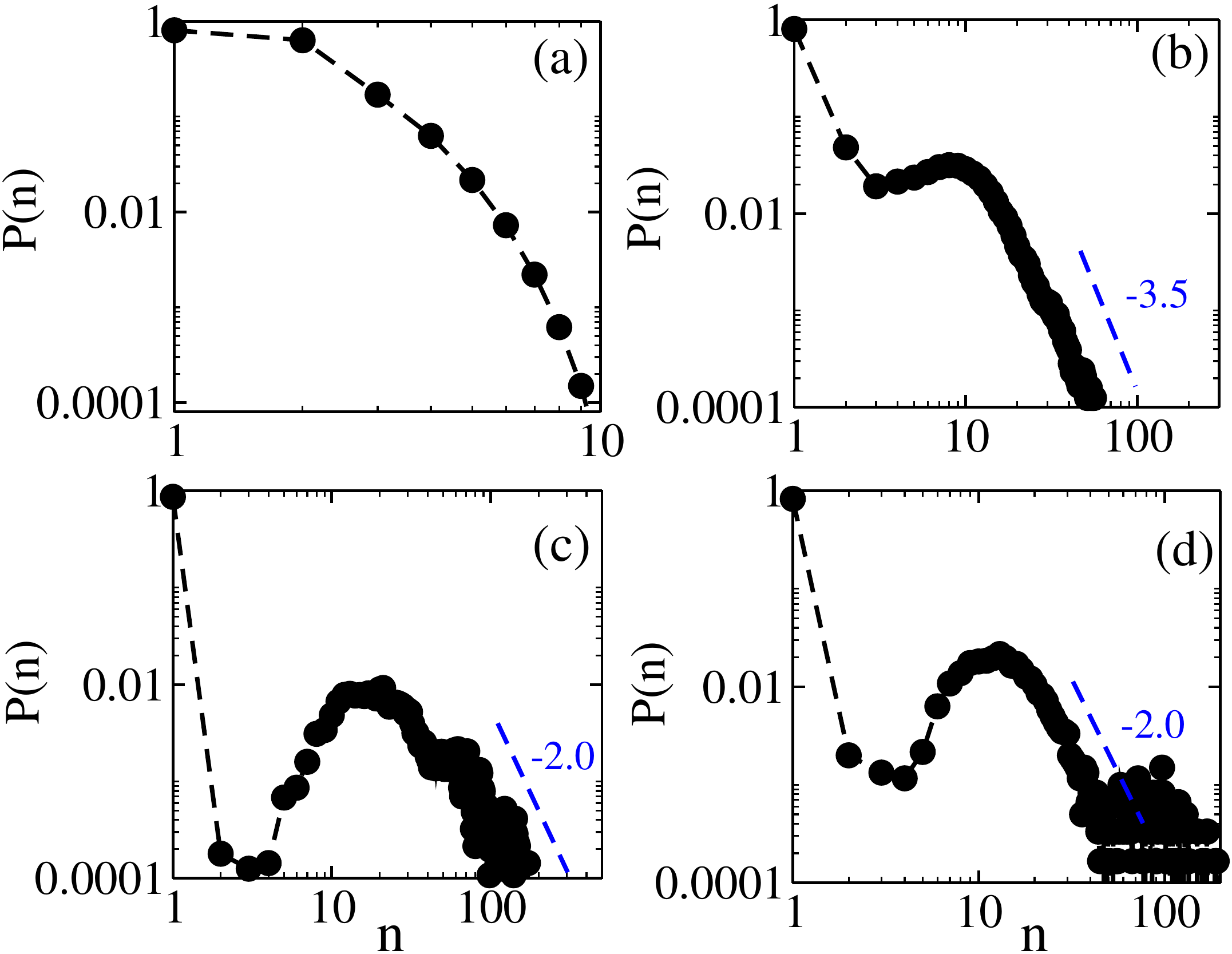}
	 \caption{(color online) We plot cluster size distribution CSD $P(n)$ vs. mean number of particles $n$ for four distinct structures. In panel (a): for Disorder structure $(D)$, we show the CSD for fixed S=4 and activity $\bar{V}=160$. (b) Disorder cluster $(DC)$: for fixed $\bar{V}=160$ and size ratio S=5. (d) Ordered Cluster $(OC)$: $\bar{V}=160$ and S=10. (c) Poly-crystalline structure $(PC)$: for fixed $\bar{V}=40$ and size ratio S=10. In disorder structure, CSD decays exponentially while polycrystalline and order cluster, CSD decay with power law with exponent -2.0. Further, for disorder cluster, CSD decays with power of exponent -3.5. Number of passive particles $N_p=400$.}
\label{fig:fig4}
\end{figure}
\begin{figure}[ht]
\centering
\includegraphics[width=1.0\linewidth]{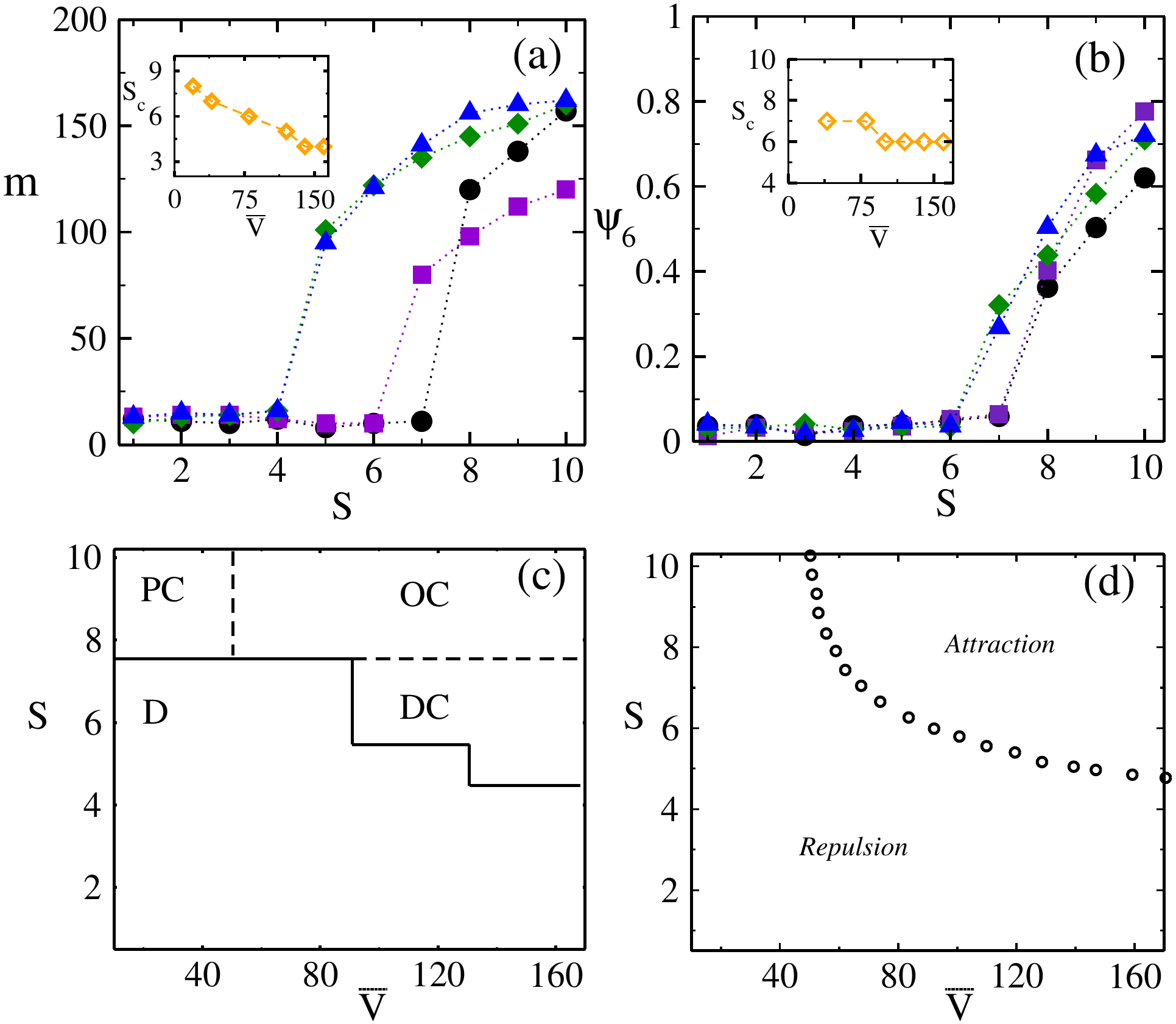}
	 \caption{(color online) 
In panel (a), we show mean cluster size $m$ vs. S, where black $\circ '^s$, voilet $\square '^s$, green $\diamond '^s$, and blue $\triangle '^s$ represent different $\bar{V}= 40,80,120,$ and 160 respectively. Inset:(a) shows the variation of critical size $S_c$ vrs. $\bar{V}$ with respect to mean size m. Panel (b) represent variation of $\psi_6$ vrs. S, where black $\circ '^s$, brown $\square '^s$, green $\diamond '^s$, and blue $\triangle '^s$ represent different $\bar{V}= 40,80,120,$ and 160 respectively. Inset:(b) shows the variation of critical size $S_c$ vrs. $\bar{V}$ with respect to mean hexatic order parameter $\psi_6$.
(c) In this panel we show the full phase diagram of four different structures in S-$\bar{V}$ plane, taking care the restrictions over $\psi_6$ and m. Dashed lines show the phase coexistent between two phases. (d) Show the nature of effective interaction in $S-\bar{V}$ plane. In all the plots number of passive particles are 400.}
\label{fig:fig5}
\end{figure}
\subsection{Steady state structural cross-over  \label{FDP}}
We extract the steady state structure of the {\em pure} passive particles with $V^{eff}(r)$. Note that the active bath particles are not explicitly considered in these simulations.  We show a representative particle snapshot on the left-hand panels for structurally distinct configurations. The radial correlation functions, $g_2(r)$, given by the distribution of the pair separations between the particles in the system, are shown on the right-hand side panel corresponding to the snapshot. We find the following four structures of passive particles.\\ 
(1) {\em Disordered structure} $(D)$: The passive particles are  homogeneously distributed for $(S,\bar{V})=(4,160)$ as shown in Fig.\ref{fig:fig3}(a). Fig.\ref{fig:fig3}(b) shows  the radial correlation function $g_2(r)$ vs. $\frac{r}{r_p}$ . The $g_2(r)$ shows a single peak at the diameter of the particle and then decay monotonically at larger distances, confirming the disordered structure.

(2) {\em  Disordered clusters} $(DC)$: The passive particles form big clusters, 
but the particles are distributed randomly within the cluster, as  shown in the representative snapshot in Fig.\ref{fig:fig3} (c) for $(S,\bar{V})=(5,160)$.
The right plot Fig. \ref{fig:fig3}(d) that  $g_2(r)$ bears signature of short ranged positional order. 

(3) {\em  Ordered cluster} $(OC)$: The particles are arranged in big clusters with local hexagonal order for $(S,\bar{V})=(10,160)$ as shown in Fig.\ref{fig:fig3}(e). The $g_2(r)$ data shows the strong periodic peaks. The location of second and third peaks appear at $\sqrt {3}$ and $2$ times the location of first peak as shown in Fig.\ref{fig:fig3}(f), consistent with the hexagonal packing.

(4) {\em Poly-crystalline structure} $(PC)$: The passive particles form ordered  domains of different mutual orientations,  shown in the snapshot of Fig. \ref{fig:fig3}(g) for ($S, \bar{V})=(10,40)$. The $g_2(r)$ plot in Fig. \ref{fig:fig3}(h) shows spilt first peak and rather broad higher peaks, suggesting the presence of more than one structure. Although there is some periodicity present as shown in the snapshot on left Fig. \ref{fig:fig3}(g) and location of different peaks in $g_2(r)$ in Fig. \ref{fig:fig3}(h).

Next we quantify the cluster size distribution in the system for different structures in the steady state. A cluster is defined as a set of particles connected by a most probable distance $r_0$. Here we choose the position of the first peak of $g_2(r)$ of the disorder structure as $r_0$. We define  the fraction of cluster of size $n$ as the cluster size distribution $(CSD)$ $P(n)$. The normalised  $P(n)$ for different  phases are shown in Fig. \ref{fig:fig4} (a)-(d)in {\it{log-log}} scale. In the $D$ structure, $P(n)$ in Fig. \ref{fig:fig4}(a) show the small clusters. For $DC$ structure in Fig.\ref{fig:fig4}(b), $P(n)$ shows an additional peak around for finite size, following which  there is  a steep decay $n^{-3.5}$  for large n. The peak for larger $n$ gets prominent for  $OC$ and $PC$, shown in Fig.\ref{fig:fig4}(c) and (d) respectively. The tail  of $P(n)$ decay less steeply than $DC$ phase, with power law exponent $-2.0$. 

We characterise the steady state structures employing: (1) the size of the largest cluster $m$  and (2) the bond orientation order parameter $\psi_6$ \cite{Mermin, Lech}. 
In $2D$ the bond orientation order parameter $\psi_6$ is defined as:
\begin{equation}
\psi_6=\frac{1}{N_p}\sum^{N_p}_{k=1}\sqrt{{\frac{1}{N_k}}\sum^{N_k}_{j=1}e^{i6\theta_{kj}}}	
\label{eq(4)}
\end{equation}
where $N_p$ is the total number of passive particles and $N_k$ shows the number of particles in the neighbour of $j^{th}$ particle. $\theta_{kj}$ is the angle between the bond connecting the $k^{th}$ and $j^{th}$ particles concerning the x-axis.  $\psi_6\sim 0$ and $\psi_6\sim 1.0$ describe the disordered and perfect hexagonal packed structure respectively.
The values of $m$ and $\psi_6$ are  shown in Table \ref{table:table1} for cases  in Fig.\ref{fig:fig3}. The error bars in the number shows the range of $m$ and $\psi_6$ for different $S$ and $\bar{V}$  where the similar structures are found.
The $m$ and $\psi_6$ values suggest that  $PC$ have cluster size and orientation order in between $OC$ and $DC$ and are not structurally distinct. 

\begin{table}[ht] 
  \caption{Characterisation of structures with respect to $m$ and $\psi_6$} 
\label{table:table1}     
\begin{tabular}{ |p{1.70cm}|p{1.70cm}|p{1.7cm}|}
\hline
$Structures$&$m$&$\psi_6$ \\ \hline
$D$&8.0$\pm 5$ & 0.05$\pm 0.012$ \\ \hline
$DC$&75.0$\pm 20$ & 0.3$\pm 0.120$ \\ \hline
$OC$&170.0$\pm 25$ & 0.70$\pm 0.100$ \\ \hline  
$PC$&138.0$\pm 20$& 0.63$\pm 0.15$ \\ \hline 
\end{tabular}\\
\end{table}

In Fig.\ref{fig:fig5} (a) we show the variation of $m$ with S for different $\bar{V}$. We find that $m$ shows a jump to large values beyond a $S_c$, a critical value of $S$. The inset shows that $S_c$  decreases linearly with $\bar{V}$. This suggests that larger $\bar{V}$ favours the formation of larger clusters. We show in Fig. \ref{fig:fig5}(b) the variation of $\psi_6$ as a function of $S$ for different $\bar{V}$. We observe that a long-ranged crystalline order is set up above $S_c$ for different $\bar{V}$ with a small jump in the order parameter value. The inset shows that $S_c$ based on $\psi_6$ does not show strong sensitivity on $\bar{V}$, unlike that determined from the magnitude of $m$. This suggests that the formation of a large cluster is sensitive to $\bar{V}$, but the orientation order is primarily sensitive to $S$.

Next, we consider the full steady state structural cross-over diagram to approximately demarcate the boundaries in $S-\bar{V}$ plane as shown in Fig. \ref{fig:fig5}(c) based on the values of m and $\psi_6$. The cross-over diagram is divided broadly into two regions by the solid line where region $D$ represents the disordered region, while the region above the solid line shows different clustered regions ($OC$ and $DC$) divided by dashed lines. The $D$ region is characterised by small $m$ along with small $\psi_6$. The $DC$ structure corresponds to large $m(~60)$ but small $\psi_6$, while the $OC$ structures correspond to large values of both. The disordered structure crosses over to clusters for sufficiently large $S$ for a given  $\bar{V}$. the boundary shifts to lower $S$ with increasing $\bar{V}$ which is consistent with the data in the inset of Fig.\ref{fig:fig5}(a). The disordered clusters get ordered one where the boundary is independent of (${S}$) as observed in the inset of Fig.\ref{fig:fig5}(b).  On the other hand, for large $S$ and low $\bar{V}$, the disordered structure crosses over to poly-crystalline $(PC)$ domains.

It may be interesting to correlate the cross-over boundaries to the changes in the nature of $V^{eff}(r)$. The boundary between repulsive and attractive $V^{eff}(r)$ is shown in Fig.\ref{fig:fig5}(d). $V^{eff}(r)$ is repulsive for low $\bar{V}$. However, for larger $\bar{V}$, there is a cross-over from repulsive interaction for low $S$ to attractive interaction for larger $S$. The disordered structure is favoured in the steady state for effective repulsion between the passive particles, while the clusters are favoured when the interaction is attractive. Large $\bar{V}$ means that the larger persistence length $l=v \tau$ or the  smaller persistence time $\tau$. Hence, the ABPs undergo a large number of collisions while the passive particles change mutual separation. This results in a scenario not too different from the equilibrium counterpart. 
In analogy to the depletion mediated attraction\cite{Harder1,Angel1}, an effective attraction occurs between the passive particles for large size differences.

The cross-over from disordered structure to the poly-crystalline domains takes place even if the interaction remains repulsive in the low $\bar{V}$ regime. In this regime, $\tau$ is large so that the separation variable of the passive particles as a dynamical variable is more strongly coupled to the dynamics of the ABPs. Both the strength and the range of repulsive interaction increase with $S$ and, hence larger effective  Barker-Henderson hardcore diameter \cite{Yiping}.
This leads to better packing among the passive particles which leads to partial orientation order in the system. The orientation order in this regime is purely a steady state effect, for the system parameters are far from order formation in equilibrium.

\begin{figure}[ht]
\centering
\includegraphics[width=0.930\linewidth]{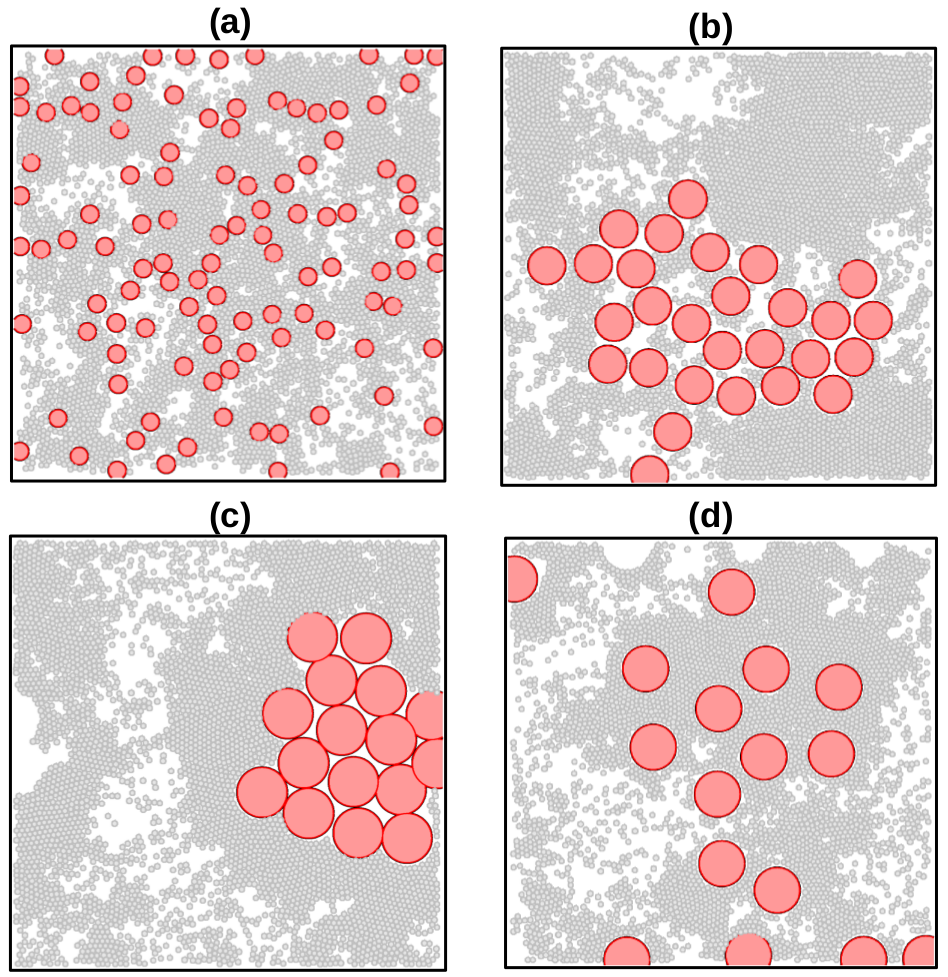}
	 \caption{(color online) Plot (a)-(d) show the four distinct structures obtained from microscopic simulation of ABPs and passive mixture with packing fraction $\phi_a=0.5$ and $\phi_p=0.2$. The parameters $\bar{V}$ and S are randomly chosen from the phase diagram for four structures shown in Fig.\ref{fig:fig5}(c) . (a) represents the disorder structure $(D)$  for $(S,\bar{V})=(4,160)$. (b) shows the disorder cluster $(DC)$ for $(S,\bar{V})=(7,160)$. (c) shows the order cluster $(OC)$ for $(S,\bar{V})=(10,160)$. (d) represents poly-crystalline structure $(PC)$ for $(S,\bar{V})=(9,40)$.  Smaller gray particles are ABPs, and red are passive ones.}
\label{fig:fig6}
\end{figure} 
\subsection{Full microscopic simulations}
Further, the results obtained in the coarse-grain simulation are confirmed by the full microscopic simulations of a mixture of active and passive particles with $\phi_a=0.5$ and $\phi_p=0.2$. In these simulations, we introduce the full microscopic interaction between the active and passive particles. Furthermore, the dimension has been taken as $L_x=L_y=800r_a$ with periodic boundary conditions in both directions. The position and orientation updates of ABP is given by Eq.\ref{eq(1)} and \ref{eq(2)}, and passive particles are updated using the following equation
\begin{equation}
	\partial_t{\bf{r}}_i^{p}=\mu_2\sum_{j\neq i}{\bf {F}}_{ij}
	\label{eq:eq5}
\end{equation}
Other simulation details are as discussed in the model section. The system is simulated for total time steps 
of $t=10^7$. The steady state structures of passive and active particles are observed in the steady state for different
size ratios $S$ and activity $\bar{V}$.
We show in Fig.\ref{fig:fig6}  four structures obtained by full microscopic simulation of passive particles in the sea of the active particles. In Fig.\ref{fig:fig6} (a), (b), (c), and (d) represent the $D$, $DC$, $OC$ and $PC$ for the parameters $S,\bar{V}=(4,160),(7,160),(10,160)$, and $(9,40)$ respectively. The snapshots closely resemble the four structures of a purely passive system in the coarse-grained simulation. Hence, results obtained from the coarse-grained simulation of purely passive particles mixture are consistent with the results obtained for the full microscopic simulation of a binary mixture of active and passive particles.

\section{conclusion \label{conclusion}}
We have studied a reduced model for steady-state structural cross-over of large passive particles in a bath of small Brownian active 
particles in two dimensions using the Langevin dynamics simulations. The effect of the active particle bath is taken into account through the effective potential between the passive particles.
The activity $\bar{V}$ and the size ratio $S$ are the two main control parameters in the system.  We observe four different steady state structures of passive particles, 
namely $D$, $DC$, $OC$, and $PC$ distinguished by the largest cluster size and the bond orientation order parameter.  Finally, the full microscopic simulations for the binary mixture of active and passive particles 
reproduce the four structures. This shows that the single component effective potential reproduces the structural features. Our study can be useful to understand the collective behaviour of passive particles in active baths,
for example, crystallisation of passive colloids, segregation of protein, bacterial suspensions, cell
suspensions, paint industry, and so on. It will be interesting to study the dynamics of the effective single component system to arrive at a comprehensive understanding of a passive system in an active bath.

\section{Acknowledgments\label{Acknowledgment}}
J.P Singh and S. Mishra thank the support and
the resources provided by PARAM Shivay Facility under the National Supercomputing Mission, Government of India at
the Indian Institute of Technology, Varanasi are gratefully
acknowledged. The computing facility at Indian Institute of
Technology (BHU), Varanasi is gratefully acknowledged.


\end{document}